\documentclass{article}
\usepackage{spconf,amsmath,graphicx,wasysym, MnSymbol}
\usepackage{hyperref}
\usepackage{bm}

\title{Structural Connectome Validation Using Pairwise Classification}
\name{Dmitry Petrov$^{\star \dagger}$\thanks{For discussion please contact  \href{mailto:to.dmitry.petrov@gmail.com}{to.dmitry.petrov@gmail.com}} \qquad Boris Gutman$^{\dagger}$ \qquad Alexander Ivanov$^{\star}$ \qquad Joshua Faskowitz$^{\dagger}$} 

\secondlinename{Neda Jahanshad$^{\dagger}$ \qquad Mikhail Belyaev$^{\ddagger \star}$\qquad Paul Thompson$^{\dagger}$}

 \address{$^{\star}$Kharkevich Institute for Information Transmission Problems, Moscow, Russia \\
     $^{\dagger}$Imaging Genetics Center, University of Southern California, Los Angeles, USA \\
     $^{\ddagger}$Skolkovo Institute of Science and Technology, Moscow, Russia}

\begin{document}
%
\maketitle
\begin{abstract}

In this work, we study the extent to which structural connectomes and topological derivative measures are unique to individual changes within human brains. To do so, we classify structural connectome pairs from two large longitudinal datasets as either belonging to the same individual or not. Our data is comprised of 227 individuals from the Alzheimer’s Disease Neuroimaging Initiative (ADNI) and 226 from the Parkinson's Progression Markers Initiative (PPMI). We achieve 0.99 area under the ROC curve score for features which represent either weights or network structure of the connectomes (node degrees, PageRank and local efficiency). Our approach may be useful for eliminating noisy features as a preprocessing step in brain aging studies and early diagnosis classification problems.


\end{abstract}
\section{Introduction}
\label{sec:intro}

Predictive modeling of neurodegenerative diseases using diffusion MR-based structural connectomes has become a popular sub-genre of neuroimaging \cite{arbabshirani2016single}. The great variety of possible pre-processing approaches needed for connectome construction leads to potential challenges in downstream application of the connectomes, for example, in a classification task. Choices of e.g, non-linear registration, parcellation, or tractography, may all have a substantial impact (\cite{zhan2015comparison}, \cite{de2013parcellation}). This state of affairs presents a challenge both in terms of intrinsic connectome reliability, and the degree to which the recovered connectomes are valid, if summary, representations of true brain connectivity (\cite{besson2014intra}, \cite{owen2013test}, \cite{bonilha2015reproducibility}, \cite{donahue2016using}).

At the same time, the performance of a particular case-control classifier may not suffice as a means of data verification due to small samples and high dimensionality. Alternative, more objective validation may be needed, such as the frequently used Intra-class Correlation Coefficient (ICC) on test-retest data (\cite{zuo2010reliable}, \cite{zhao2015test}). However, structural connectome ICC is generally low, which complicates method comparison. Also, the parametric constraints of classic ICC \cite{muller1994critical} may not be valid. Non-parametric approaches free of data distribution assumptions may be more suitable.

To address this issue, we propose a pairwise classification approach to intrinsically assess connectome utility across time, somewhat in line with a recent method for functional connectomes as well \cite{Finn2015}. For each set of connectomes ${C^i_j}$ and features ${f(C^i_j)}$ in question, we construct all possible pairs $(C^{i_1}_{j_1}, C^{i_2}_{j_2})$, where $i$-indices correspond to images and $j$-indices correspond to subjects. We then conduct a linear classification on the pairwise differences $\|f(C^{i_1}_{j_1}) - f(C^{i_2}_{j_2})\|$ of these pairs with respect to the target variable $y$: $y=1$ if~$j_1 = j_2$, $0$ else.

We test this pipeline on structural connectomes derived from two publicly available neuroimaging datasets: ADNI and PPMI. These datasets have scanned subjects multiple times, with at least a one year interval between scans. We~achieve 0.99 ROC AUC both for direct connectome measures (bag of edges) and for features representing connectome structure (PageRank), suggesting that the tested data is reliable enough to distinguish subjects by the proposed approach. Similar research was conducted by Yeh et al. \cite{yeh2016quantifying}. Though the authors used a local structural connectome, different features, datasets and connectome construction pipelines, they arrived at similar conclusions.

\section{Pairwise classification}

We propose the following pipeline for pairwise connectome classification: normalization, building connectome features, building pairwise features based on connectome features. Let's denote a set of connectomes as $\{C^i_j\}$, where $j$ is an index of a subject and $i$ is an index of an image. 

\subsection{Normalizations}

Topological normalization of connectivity matrices may be useful prior to any analysis, because the number of detected streamlines is known to vary from individual to individual and can also be affected by fiber tract length, volume of cortical regions and other factors (\cite{hagmann2007mapping}, \cite{bassett2011conserved}). There is no consensus on the best normalization approach, so we use the three following topological normalization schemes alongside with pure weights (no normalization at all) -- by mean, by maximum and binary normalization with zero threshold: $
a^b_{kl}~=~1~\text{if}~a_{kl} > 0, 0~\text{else}$ where $a_{kl}$ is a connectome edge.




\subsection{Network features}

For each connectome and each normalization we build ``bag of edges" vectors from the upper triangle of the symmetric connectivity matrix. In addition, we calculate eight network metrics for each node: weighted degrees, or strength; closeness, betweenness and eigenvector centralities; local efficiency; clustering coefficient; weighted number of triangles around node. We choose these features because they are well-described and reflect different structural properties of connectomes \cite{rubinov2010complex}. We also calculate PageRank for each node. Introduced in 1998 by Brin and Page \cite{pagerank_original} this metric roughly estimates probability that a person randomly clicking on links in the network will arrive at particular node.

\subsection{Pairwise features}
\label{pairs}

Each normalization and set of features described above defines a mapping from connectome space to feature space $C \rightarrow f(C)$. Since our goal is to check how well this mapping separates connectomes in it, we propose various pairwise features. For each set of connectome features in question we make all possible pairs of connectome features -- ($f(C^{i_1}_{j_1}), f(C^{i_2}_{j_2})$). For each pair, we assign a binary target variable -- 1 if connectomes are from the same subject ($j_1 = j_2$), 0 -- if they are from different subjects ($j_1 \neq j_2$). Finally, for each pair we build a vector of three features, describing their difference $\|f(C_1) - f(C_2) \|$ according to $l_1$, $l_2$ and $l_{\infty}$ norms.

\subsection{Classifiers and validation}

We use linear classifiers for pairwise classification: logistic regression (LR), SVM with linear kernel and stochastic gradient descent (SGD) with modified Huber loss. We scale features with standard scaling and apply elastic-net regularization for each of classifiers.

Model performance we measure with area under ROC curve (ROC AUC) through a two-step validation procedure. First, for each dataset, we perform hyperparameter grid search based on a 10-fold cross-validation with a fixed random state for reproducibility. For each model we varied overall regularization parameter, $l_1$-ratio and number of iterations for SGD. Then we evaluate the best parameters on 100 train/test splits with fixed different random states (test size was set to 20\% of data). We report the ROC AUC distribution on these 100 test splits for each combination of normalization/base features/diagnostic group in results section.

\section{Experiments}

\subsection{Base data}

We used two datasets for our experiments. Our first dataset, the Alzheimer’s Disease Neuroimaging Initiative (ADNI2), is comprised of 227 individuals (675 scans), mean age at baseline visit $73.1 \pm 7.4$, 99 females. Each individual had at least 1 brain scan and at most 6 scans. The data include 46 people with AD (111 AD scans), 80 individuals with EMCI (247 MCI scans), 40 people with LMCI (120 LMCI scans) and 61 healthy participants (160 scans). Second, we used imaging data from the Parkinson’s Progression Markers Initiative (PPMI) database. From it we selected subjects with PD (159 subjects) and healthy controls (67 subjects). These included a total of 226 individuals (456 scans), mean age at the baseline $61.0 \pm 9.8$ years, 79 females. Each individual had at least 1 brain scan and at most 4 scans.

\subsection{Network construction}

Inhomogeneity corrected T1-weighted images for ADNI and PPMI data were processed with FreeSufer's \cite{fischl2012freesurfer} recon-all pipeline to obtain a triangular mesh of the gray-white matter boundary registered to a shared spherical space, as well as corresponding vertex labels per subject for a cortical parcellation based on the Desikan-Killiany (DK) atlas\cite{desikan2006automated}. This atlas includes 68 cortical brain regions; hence, our cortical connectivity matrices were 68$\times$68.

In parallel, T1w images were aligned (6-dof) to the 2mm isotropic MNI 152 template. These were used as the template to register the average $b_0$ of the DWI images, in order to account for EPI related susceptibility artifacts. DWI images were also corrected for eddy current and motion related distortions. Rotation of b-vectors was performed accordingly. Tractography for ADNI data was then conducted using the distortion corrected DWI in 2mm isotropic MNI 152 space. Probabilistic streamline tractography was performed using the Dipy \cite{garyfallidis2014dipy} LocalTracking module and implementation of constrained spherical deconvolution (CSD) \cite{tax2014recursive} with a spherical harmonics order of~6. Streamlines longer than 5mm with both ends intersecting the cortical surface were retained. Edge weights in the original matrices are proportional to the number of streamlines detected by the algorithm.

PPMI data were processed in a slightly different fashion to account for variability in the acquisition protocols and to show our method is not dependent on any single processing scheme. Images were initially denoised with an adaptive denoising algorithm \cite{Denoise} and two DWI acquisitions from each subject were merged. DWI images were corrected for eddy current and motion related distortions, then non-linearly epi-corrected with ANTs SyN. Rotation of b-vectors was performed accordingly. Tractography for PPMI data was then conducted in 2mm isotropic MNI 152 space, again using the Dipy LocalTracking module. At each voxel, the CSD was fitted recursively \cite{tax2014recursive} with a spherical harmonic order of~6. Deterministic streamline tractography was seeded at two random locations in each white matter voxel. Similar to the ADNI data, only streamlines longer than 5mm with both ends intersecting the cortical surface were retained. 

\subsection{Pairwise data}

For each set of connectomes described above (ADNI, PPMI) we made all possible pairs of connectomes as described in \ref{pairs}. Using this technique we obtained 227475 pairs (764 of which were labeled as 1) from ADNI2 data and 152031 pairs from PPMI data (301 of which were labeled as 1). Due to huge imbalance of classes in generated pairs, we used all samples with label 1 and equally sized random subsample of 0. Our result do not depend for different subsamples of 0s, so we report them for a fixed random state.

\subsection{Pairwise classification performance}

\begin{figure}[htb]
\begin{minipage}[b]{1.0\linewidth}
  \centering
  \centerline{\includegraphics[width=8cm]{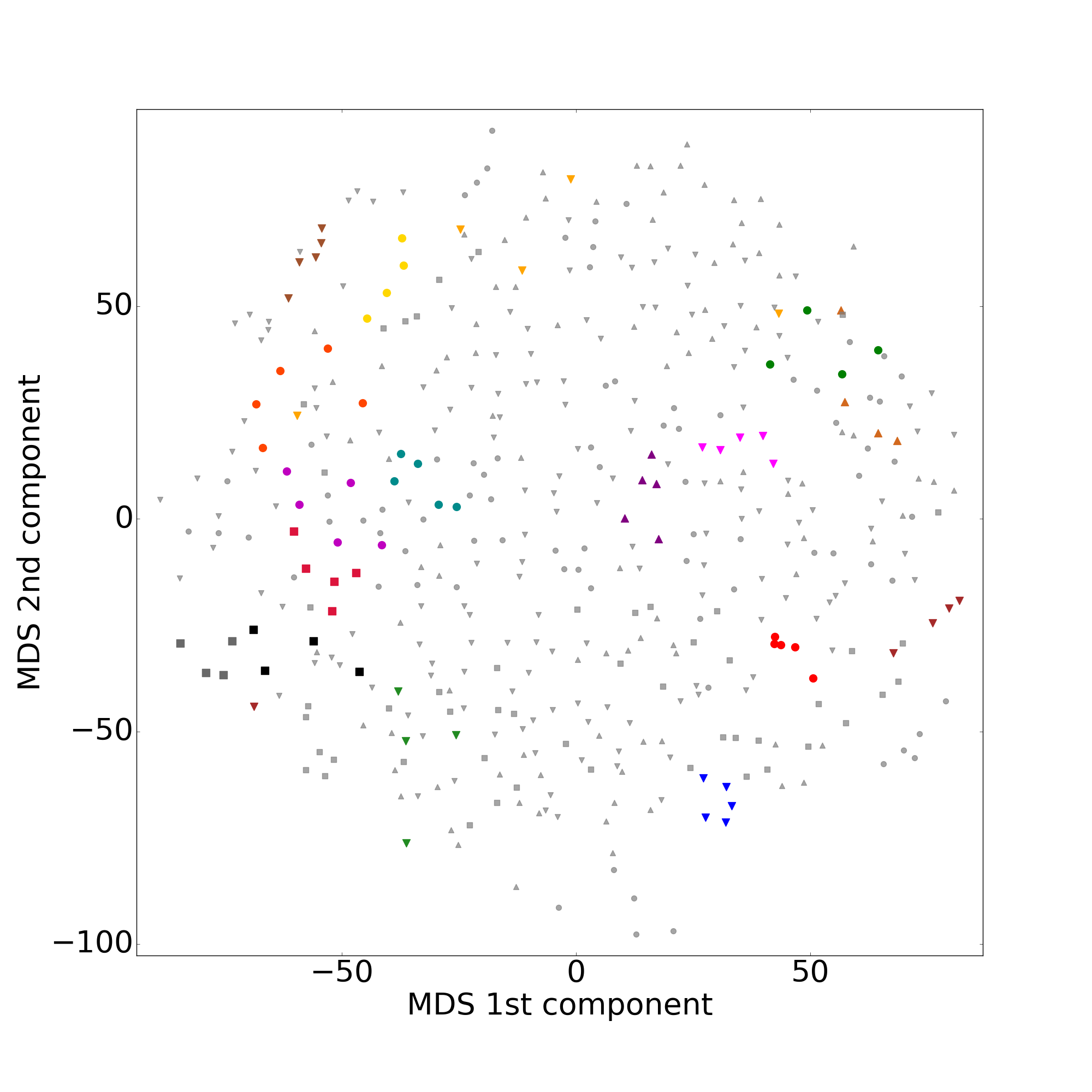}}
  \caption{Multidimensional scaling for ADNI data based on $l_2$~norm. Points with same colors represent same subjects. To avoid mess on the pictures we've decided to color only 17 subjects. Other subjects represented by smaller grey markers. Marker shape indicates diagnostic group: $\medcircle$ -- controls, $\medtriangleup$ -- LMCI, $\medtriangledown$ -- EMCI, $\medsquare$ -- AD. }\medskip
\end{minipage}
\end{figure}

Figure 1 shows multidimensional scaling (MDS) based on $l_2$-norm dissimilarity matrix of bag of edges for ADNI subjects (for PPMI data picture is essentially the same, so we omitted it). We see that in most cases images from same subject are near each other in that feature space. We also see that there is no such clear picture with diagnostic groups labels.

\begin{figure}[htb]

\begin{minipage}[b]{1.0\linewidth}
  \centering
  \centerline{\includegraphics[width=10cm]{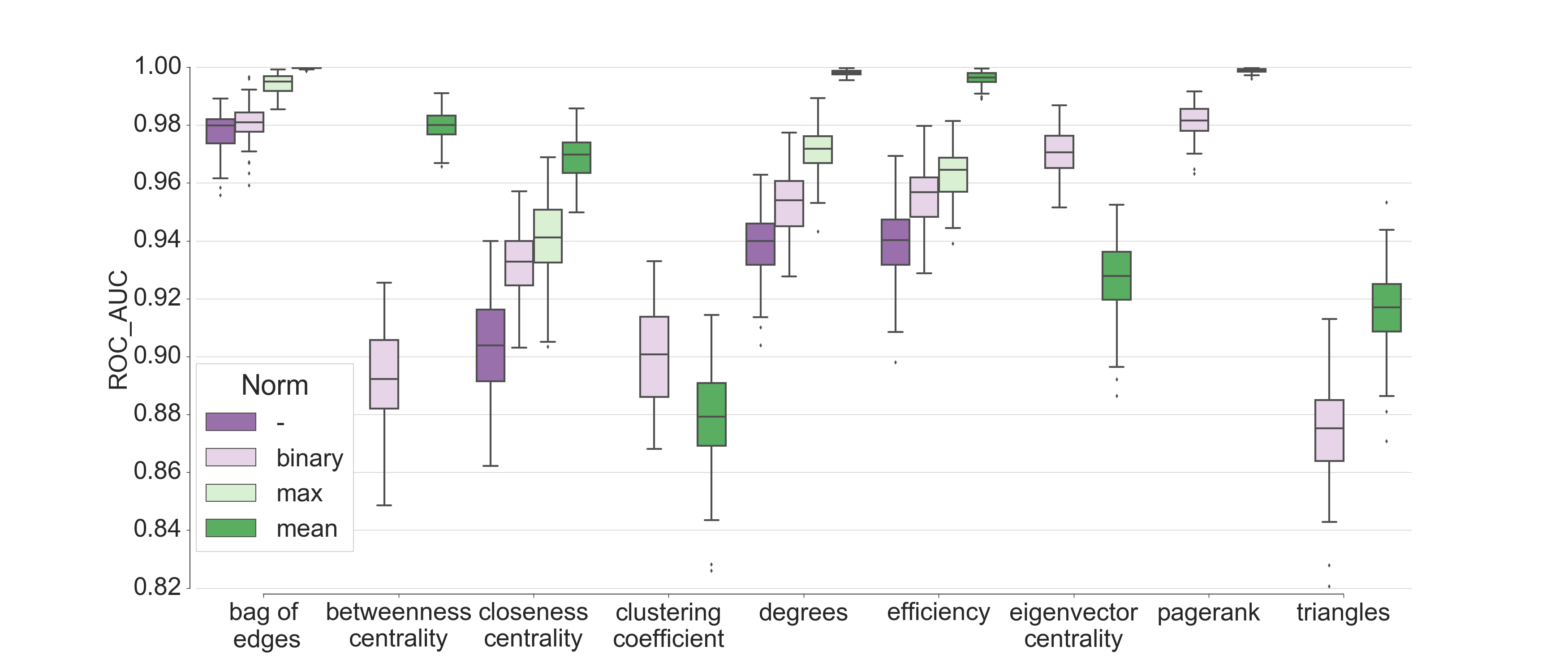}}
  \caption{ROC AUC distributions for pairwise classification on all ADNI data depending on choice of connectome normalizing scheme and base features. For betweenness centrality, clustering coefficient, eigenvector centrality, PageRank and triangles, we report results only for the binary and mean normalizations. Normalizing by the other norms affects these features identically to the mean normalization.}\medskip
\end{minipage}
\end{figure}

\begin{figure}[htb]
\begin{minipage}[b]{1.0\linewidth}
  \centering
  \centerline{\includegraphics[width=10cm]{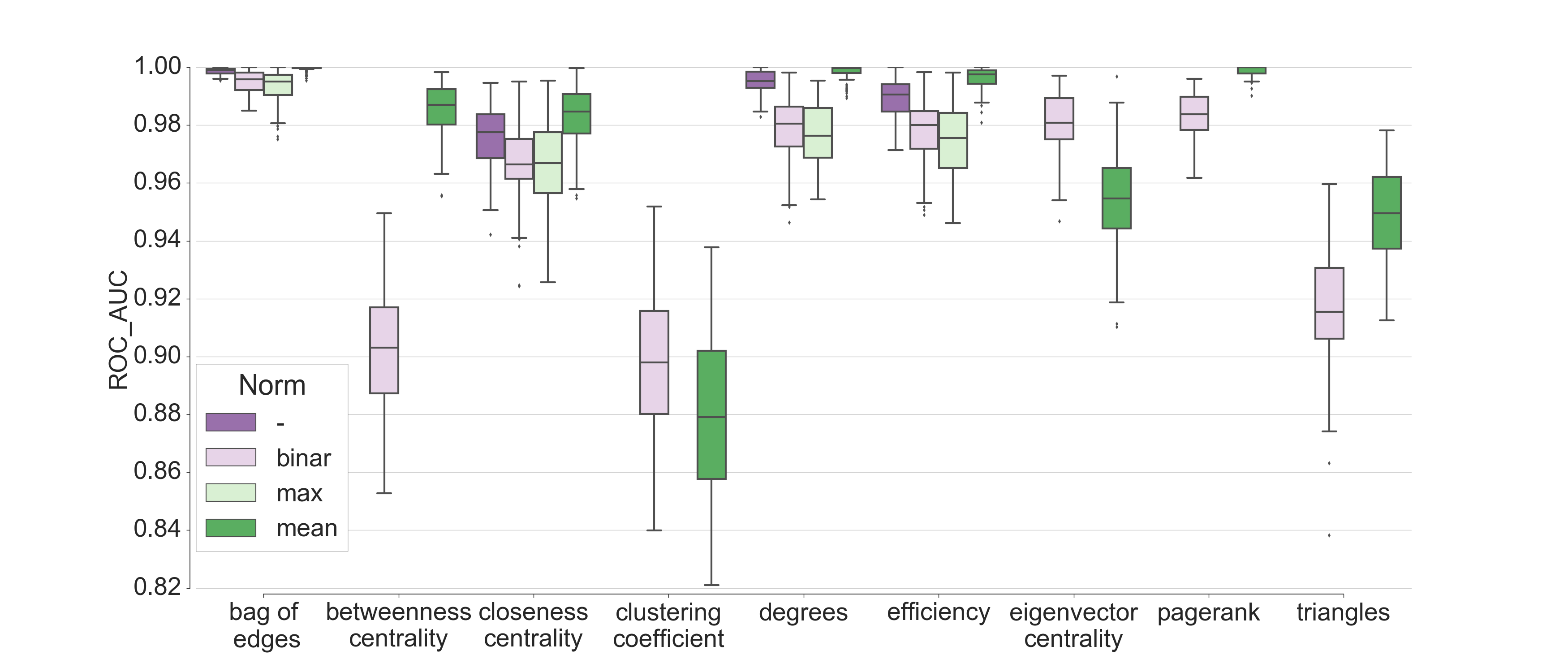}}
  \caption{ROC AUC distributions for pairwise classification on all PPMI data depending on choice of connectome normalizing scheme and base features. For betweenness centrality, clustering coefficient, eigenvector centrality, PageRank and triangles, we report results only for the binary and mean normalizations. Normalizing by the other norms affects these features identically to the mean normalization.}\medskip
\end{minipage}
\end{figure}

\begin{figure}[htb]
\begin{minipage}[b]{1.0\linewidth}
  \centering
  \centerline{\includegraphics[width=10cm]{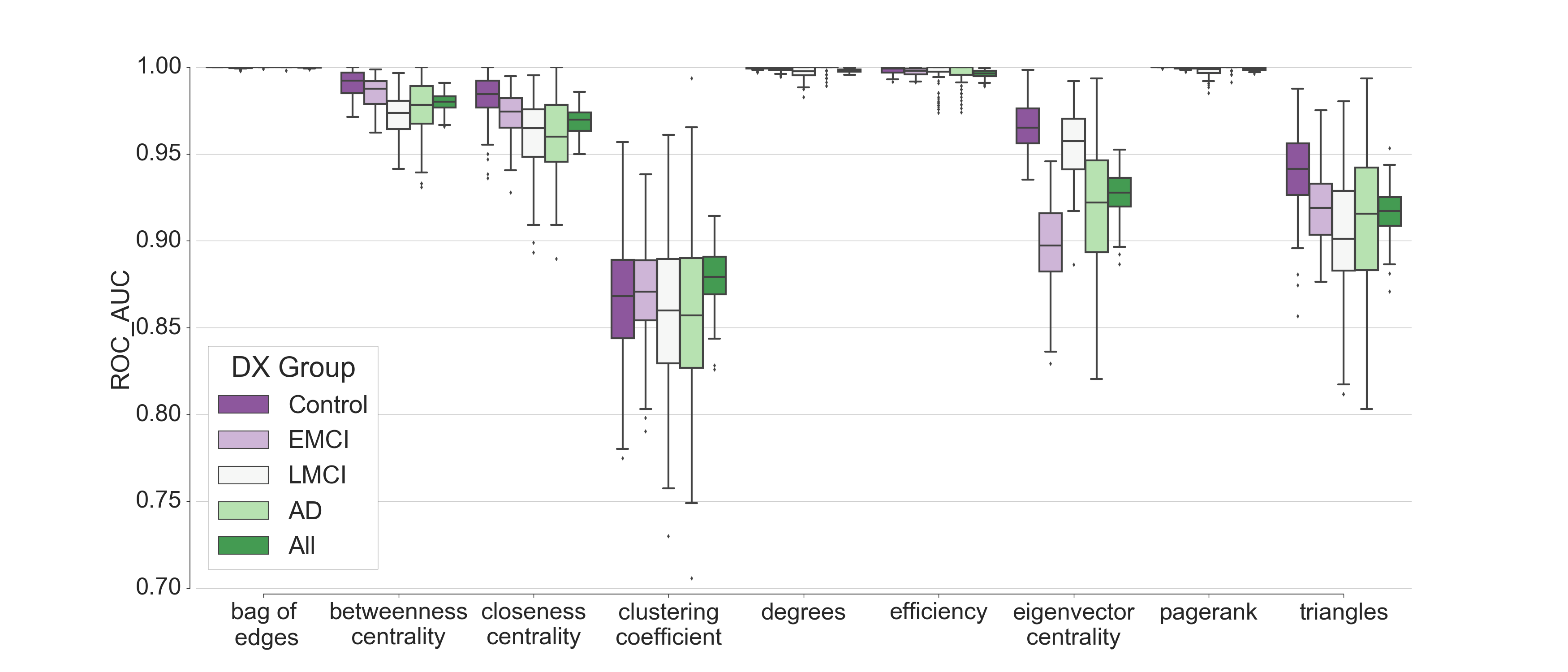}}
  \caption{ROC AUC distributions for pairwise classification on ADNI data for different diagnostic groups on different ``base" features. These results are reported for connectome normalization by mean.}\medskip
\end{minipage}
\end{figure}

\begin{figure}[htb]
\begin{minipage}[b]{1.0\linewidth}
  \centering
  \centerline{\includegraphics[width=10cm]{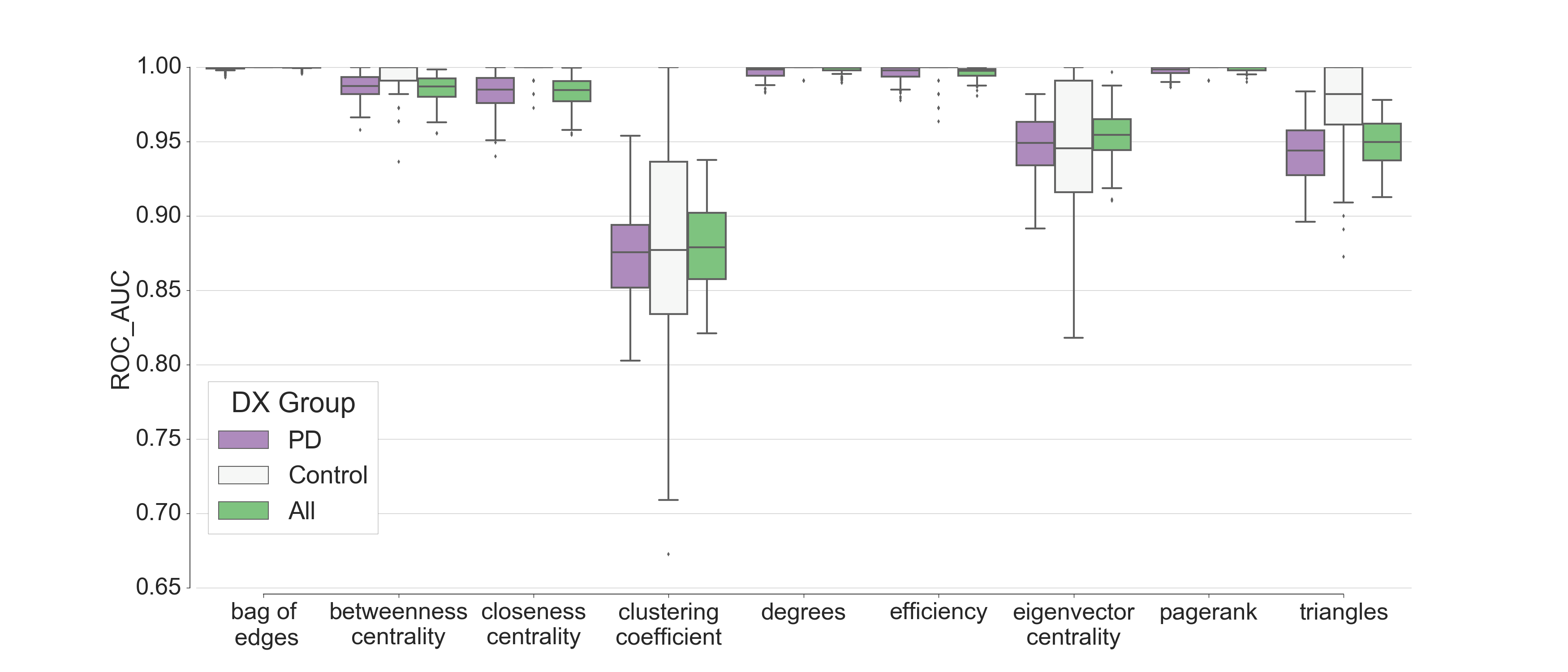}}
  \caption{ROC AUC distributions for pairwise classification on PPMI data for different diagnostic groups on different ``base" features. These results are reported for connectome normalization by the mean.}\medskip
\end{minipage}
\end{figure}

Figures 2 and 3 quantify this observation for ADNI and PPMI data in terms of ROC AUC distributions depending on normalization and base features. We see that 0.99 ROC AUC can be achieved either for connectome weights themselves, or for features that capture connectome structure. We also see that the choice of normalization greatly affects the accuracy of pairwise classification. Normalizing by the mean is a winner in most cases, with the exception of eigenvector centrality and clustering coefficient features. It is interesting to note that for clustering coefficient and eigenvector centrality, binary normalization performed better than other normalizations even though it preserves somewhat less information.

Figures 3-4 show the ROC AUC distributions of pairwise classification depending on the base features and diagnostic group for connectomes normalized by the mean. We see that there is almost no difference in pairwise classification results in different diagnostic groups. We note that interquartile spread is high for diagnostic groups with the smallest number of subjects.

\section{Conclusion}


We have presented a method for structural connectome feature validation through pairwise classification which is free of distribution assumptions. We tested this pipeline on ADNI and PPMI data and obtained high classification performance in terms of ROC AUC suggesting that there are mappings from connectomes to feature spaces that at least differentiate subjects from each other. 

It is worth noting that pairwise classification is not a feature selection technique for classification tasks. It is possible that a feature distinguish classes in the context of this work, but fails to distinguish subjects, for example in a diagnostic classification task. Our results suggest that pairwise classification may be useful for validating preprocessing pipelines and particular features in terms of how much subject-related signal they preserve. As such, it may be treated as a ``first-pass" for connectome features to be used in further studies. 

There are several limitations. First, we used only one particular pipeline to construct our networks. These results may differ for other tractography algorithms and parcellations. Assessing these effects on pairwise classification is among our future goals. 

Second, the downsampling of the ``different subject pair" class to ensure balanced samples may lead to optimistic accuracy estimates. A detailed look at the multidimensional scaling plot suggests that a number of connectomes from different subjects are ``near" each other, though it is unlikely that our downsampling procedure selects them for training or testing. To answer this question, we plan to apply our protocol without downsampling. 

Finally, due to limitations of the data used here, same-subject pairs were constructed from diffusion images acquired at least one year apart. Such a time scale provides ample opportunity for substantial longitudinal effects, such as those due to neurodegeneration, to affect our features. As we cannot exclude such effects, any conclusions to be drawn about optimal network features and normalizations for further studies must be made with the appropriate reservations.

\section{Acknowledgments}

The results of sections 2 and 3 are based on the scientific research conducted at IITP RAS and supported by the Russian Science Foundation (project 14-50-00150).

Some of data used in preparation of this article were obtained from the Alzheimer’s Disease Neuroimaging Initiative (ADNI) database. A complete listing of ADNI investigators as well as data acquisition protocols can be found at \url{adni.loni.usc.edu}. 
Additional data were obtained from the Parkinson’s Progression Markers Initiative (PPMI) database. For up-to-date information on the study, visit \url{www.ppmi-info.org}. 

\bibliographystyle{IEEEbib}
\bibliography{strings,refs}

\end{document}